\documentclass[reprint,longbibliography,aps]{revtex4-2}

\usepackage{graphicx} % Required for including pictures
\usepackage{amsmath}
\usepackage{ amssymb }

\usepackage{color}

\usepackage{float} % Allows putting an [H] in \begin{figure} to specify the exact location of the figure
\usepackage{wrapfig} % Allows in-line images such as the example fish picture

\usepackage[makeroom]{cancel} %allows strikethroughs
\usepackage{comment}

\usepackage{lipsum} % Used for inserting dummy 'Lorem ipsum' text into the template

\usepackage{outlines}

\graphicspath{{./figs/}} % Specifies the directory where pictures are stored

\newcommand{\beq}{\begin{equation}}
\newcommand{\eeq}{\end{equation}}

\newcommand{\bvec}{\begin{pmatrix}}
\newcommand{\evec}{\end{pmatrix}}
\newcommand{\lp}{\left(}
\newcommand{\rp}{\right)}

\newcommand{\llangle}{\left \langle}
\newcommand{\rrangle}{\right \rangle}

\newcommand{\ve}[1]{\mathbf{#1}}

\newcommand{\D}{\text{d}}
\newcommand{\bv}{\ve{v}}
\newcommand{\bw}{\ve{w}}

\AtBeginDocument{%
	\newwrite\bibnotes
	\def\bibnotesext{Notes.bib}
	\immediate\openout\bibnotes=\jobname\bibnotesext
	\immediate\write\bibnotes{@CONTROL{REVTEX41Control}}
	\immediate\write\bibnotes{@CONTROL{%
			apsrev42Control,author="08",editor="1",pages="1",title="0",year="1"}}
	\if@filesw
	\immediate\write\@auxout{\string\citation{apsrev42Control}}%
	\fi
}%
\begin{document}

%----------------------------------------------------------------------------------------
%	TITLE PAGE
%----------------------------------------------------------------------------------------

%\begin{titlepage}

\title{Improving the Feasibility of Economical Proton-Boron 11 Fusion via Alpha Channeling with a Hybrid Fast and Thermal Proton Scheme}

\author{Ian E. Ochs}
%\email{iochs@princeton.edu}
\affiliation{Department of Astrophysical Sciences, Princeton University, Princeton, New Jersey 08540, USA}
\author{Elijah J. Kolmes}
\affiliation{Department of Astrophysical Sciences, Princeton University, Princeton, New Jersey 08540, USA}
\author{Mikhail E. Mlodik}
\affiliation{Department of Astrophysical Sciences, Princeton University, Princeton, New Jersey 08540, USA}
\author{Tal Rubin}
\affiliation{Department of Astrophysical Sciences, Princeton University, Princeton, New Jersey 08540, USA}
\author{Nathaniel J. Fisch}
\affiliation{Department of Astrophysical Sciences, Princeton University, Princeton, New Jersey 08540, USA}

\date{\today}% It is always \today, today,

\begin{abstract}

The proton-Boron$^{11}$ (p-B11) fusion reaction is much harder to harness for commercial power than the easiest fusion reaction, namely the deuterium and tritium (DT) reaction.  The p-B11 reaction requires much higher temperatures, and, even at those higher temperatures, the cross section is much smaller.  However, as opposed to tritium, the reactants are both abundant and non-radioactive. It is also an aneutronic reaction, thus avoiding radioactivity-inducing neutrons.  Economical fusion can only result, however, if the plasma is nearly ignited; in other words if the fusion power is at least nearly equal to the power lost due to radiation and thermal conduction. Because the required temperatures are so high, ignition is thought barely possible for p-B11, with fusion power exceeding the bremsstrahlung power by only around 3\%.  We show that there is a high upside to changing the natural  flow of power in the reactor, putting more power into protons, and less into the electrons. This redirection can be done using waves, which tap the alpha particle power and redirect it into protons through alpha channeling. Using a simple power balance model, we show that such channeling could reduce the required energy confinement time for ignition by a factor of 2.6 when energy is channeled into thermal protons, and a factor of 6.9 when channeled into fast protons near the peak of the reactivity. Thus, alpha channeling could dramatically improve the feasibility of economical p-B11 fusion energy.

\end{abstract}

\maketitle
%----------------------------------------------------------------------------------------
%	TABLE OF CONTENTS
%----------------------------------------------------------------------------------------

%\widowpenalty=10000 %was supposed to do something but didnt
%\clubpenalty=10000

\section{Introduction} 
\begin{comment}

\begin{itemize}
	\item Historic focus on easier reactions
	\item Advantage of aneutronic fuels
	\item Long thought p-B11 was impossible and why (thermonuclear concerns) \cite{Rider1995,Rider1995a}
	\item Last 20 years -- interest in nonthermal concepts (Volosov\cite{Volosov2006ACT,Volosov2011Problems}, Lasers\cite{Ruhl2022LaserPB11}, beams\cite{Eliezer2020BeamPB11})
	\item Sikora and Weller\cite{Sikora2016CrossSection} -- higher cross section than previously thought
	\item Opening of regime for thermonuclear high-Q performance by Putvinski \cite{Putvinski2019}
	\item Return of interest to thermal schemes: Cai et al\cite{Cai2022TokamakPB11}--thermonuclear, but requirement for cold electrons, non-self-consistent
	\item Extreme requirements on confinement time for ignition.
	\item Altering power flow through alpha channeling can dramatically improve these requirements.
	\item Combination of thermonuclear fusion with nonthermal features can further improve these requirements.
	\item Outline of paper.
\end{itemize}

\end{comment}

Historically, fusion energy research has focused primarily on the deuterium-tritium (DT) reaction, due to its high cross section at relatively low temperature.
This feature means that the confinement requirements for achieving (DT) fusion are much lower than for other fuels, making it the most logical fuel to exploit in the near term.

However, there are several disadvantages to DT fusion.
First, tritium is radioactive. 
Second, it is not abundant, and must be bred from lithium or other materials.
Third, the DT reaction produces fast neutrons.
In addition to the proliferation risk that these entail, magnets and sensitive instruments must be shielded from these neutrons using considerable shielding material, which significantly adds to the volume and cost of any confinement device.
Over time, the neutrons break down this shielding, turning it into a structurally unsound, radioactive slab that must be safely stored away for hundreds of years.

Such deficiencies of the DT reaction have lead to an interest in aneutronic fuels.
One of the most appealing of these is the proton-Boron$^{11}$ (p-B11) reaction, which has the additional advantage of fuel abundance.

For a long time, it was thought that achieving a self-sustaining thermonuclear fusion reaction (ignition) was impossible for p-B11.
This pessimism came from the fact that the fusion cross section was too small, and occurred at too high a temperature \cite{Nevins2000CrossSection}.
Thus, it seemed that the bremsstrahlung power would always exceed the fusion power, requiring external heating power to maintain the reaction \cite{Rider1995,Rider1995a}.
This lead to a proliferation in interest in nonthermal and nonequilibrium schemes \cite{Rostoker1997BeamPB11,LampeMannheimer1998CommentsPB11,Volosov2006ACT,Volosov2011Problems,Labaune2013LaserPB11,Ruhl2022LaserPB11,Eliezer2016Avalanche,eliezer2020NovelFusion,Eliezer2020BeamPB11}, which accept the requirement for significant external heating and seek to optimize the output energy given that constraint.

Fortuitously, recent results have shown that the cross section for the p-B11 reaction is larger than previously thought \cite{Sikora2016CrossSection}.
These larger cross sections, combined with more detailed calculations of how the fusion-born alpha particles damp on the protons, have resulted in a more optimistic picture, showing that ignition is in fact possible for p-B11 in thermonuclear fusion plasmas \cite{Putvinski2019}.
This realization has led to a revival in interest in thermonuclear p-B11 fusion \cite{Cai2022TokamakPB11}.

Just because ignition is theoretically possible, however, does not mean that it is particularly feasible. 
As we show later in this paper, the ignition window identified by Putvinski \cite{Putvinski2019} requires achieving an energy confinement time of around 500 seconds at ion densities of $10^{14}$ cm$^{-3}$--an enormous technological hurdle.
Thus, it is important to examine processes which might reduce these extreme requirements.

Much of the reason for these extremely large confinement times required for ignition is that the fusion power only exceeds the bremsstrahlung power by a few percent.
Widening this gap between fusion and bremsstrahlung power to even 20\% thus has the potential to produce a 7x improvement in the required confinement time.
To do this, one must try to redirect power from the electrons (which produce radiation) to the protons (which produce fusion).

To redirect the power from the alpha particles into the protons, one can make use of waves, in a process known as alpha channeling \cite{Fisch1992,fisch1992current,fisch1995ibw}. 
This possibility was explored by Hay \cite{Hay2015Ignition}, but crucially, that paper ignored thermal conduction losses.
As can be shown by a simple analytic model \cite{kolmes2022waveSupported}, much of the aid alpha channeling provides is in dramatically decreasing the confinement time required to achieve ignition.

In this paper, we delve more deeply into examining the potential improvements to p-B11 fusion provided by alpha channeling.
We discuss the important key metrics in achieving economical fusion energy, focusing in on the importance of the minimum energy confinement time to achieve ignition $\tau_E^*$.
We then provide a simple 0D computational power balance model to evaluate this confinement time, which accounts for collisional and wave-based energy exchanges between the different species in the plasma.
As this model shows, using alpha channeling to put power directly into the protons can lower the required confinement time to achieve ignition by a factor of around 2.6.
Furthermore, alpha channeling improves the robustness of the reaction to contamination by fusion ash.

With the use of waves, however, it is no longer necessary to put the energy into thermal protons--instead, the energy can be put directly into maintaining protons near the peak of the fusion reactivity at 650 keV. 
Such a reaction can be seen as a hybrid between beam and thermonuclear fusion, as it incorporates large populations of both fast and thermal protons.
By allowing for the presence of separate fast ion population in our power balance model, we show that this hybrid scheme improves the confinement time by a further factor of three, resulting in a total factor-of-6.9 reduction in the required confinement time for ignition relative to purely thermonuclear p-B11 fusion.
These results broadly match those in \cite{kolmes2022waveSupported}, now shown with a more full optimization and a more accurate power balance model.

To demonstrate these promising results, we begin in Section~\ref{sec:performance} with a discussion of the power flow in a fusion reactor, explaining the rationale for confinement time as a performance metric for high-performance fusion plasmas, and why the p-B11 reaction is particularly challenging.
In Section~\ref{sec:powerBalanceModel}, we introduce the power balance model itself, which captures collisional exchange of energy between fast protons, thermal protons, boron, and electrons, heating by alpha particles, bremsstrahlung radiation, and alpha channeling.
In Section~\ref{sec:tauEOptimization}, we describe how to optimize the confinement time given different assumptions for the alpha channeling.
We then numerically perform this optimization, showing how alpha channeling results in much lower required confinement times for ignition.

In Section~\ref{sec:ash}, we consider the effect of poisoning by alpha particle ash, the product of the fusion reaction.
Such ash increases the bremsstrahlung power, and without alpha channeling a very small quantity of ash $(<2\%)$ can preclude ignition, even when assuming perfect confinement.
We show that alpha channeling allows for ignition at much higher ash concentrations, even when allowing for non-perfect confinement.

The core analysis of the paper is contained in Sections \ref{sec:performance}-\ref{sec:ash}.
In the subsequent sections, we briefly discuss other considerations in designing a reactor.
In Section~\ref{sec:beamTarget}, we discuss why the optimal ion mix for achieving ignition contains a mix of fast and thermal protons, rather than simply a beam of fast protons--a topic also covered in \cite{kolmes2022waveSupported}.
In Section~\ref{sec:recycling}, we briefly go over how consideration of energy recycling in the full reactor power balance might lead to even lower required confinement times. We also discuss how recycling might lead to a very different optimal mix of thermal and fast protons, if one can achieve high recycling efficiencies from direct conversion.
Finally, in Section~\ref{sec:addlPowerLoss}, we discuss additional power loss mechanisms due to the confinement systems and electron-cyclotron radiation, and how they might affect the design of a fusion reactor.

\begin{figure*}[t]
	\centering
	\includegraphics[width=0.8\linewidth]{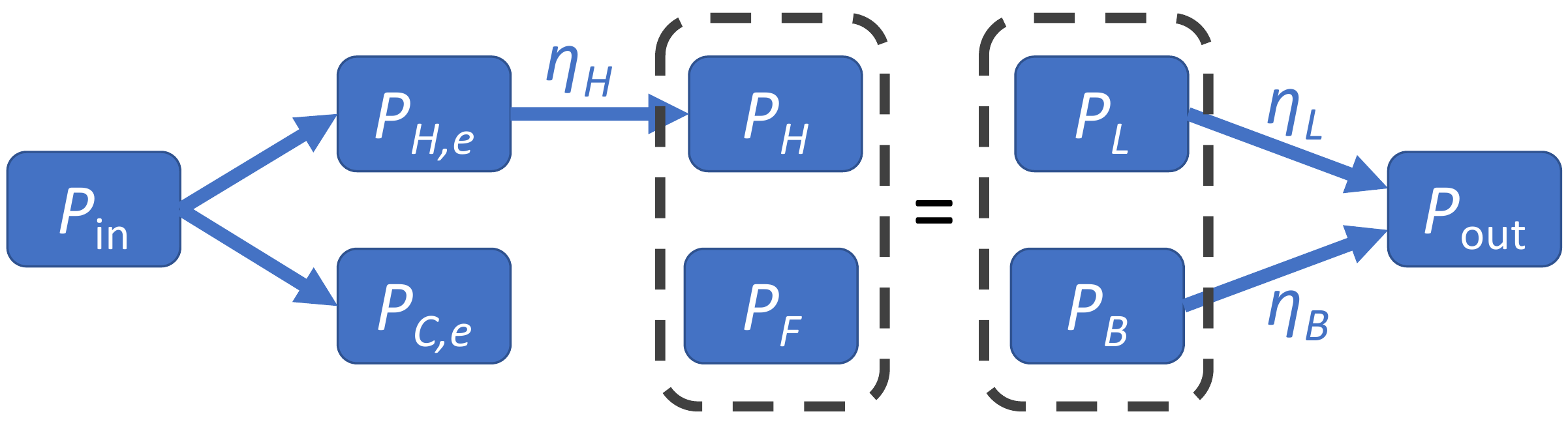}
	\caption{Simplified power flow model for a fusion power plant. Electrical power $P_\text{in}$ is split, with a portion $P_{C,e}$ supporting the confinement, and a portion $P_{H,e}$ going to heating.
	The electrical heating power is delivered with some efficiency $\eta_H$ to the plasma, resulting in $P_H$ of delivered heating power.
	This results in fusion power $P_F$.
	The hot plasma sustains power losses through thermal conduction losses $P_L$ and bremsstrahlung $P_B$, which are converted into output electrical power $P_\text{out}$ with efficiencies $\eta_L$ and $\eta_B$ respectively.
	The plasma and nuclear physics define a relationship between $P_H$, $P_F$, $P_L$, and $P_B$. 
	Engineering and technological considerations determine the various power conversion efficiencies $\eta$'s, as well as the power used for confinement $P_{C,e}$. 
	For a successful power plant, $P_\text{out} > P_\text{in}$.}
	\label{fig:reactorPowerFlow}
\end{figure*}

\section{Power Flow and Performance Metrics}\label{sec:performance}

To consider the potential advantage of altering the energy flow from the alpha particles, it is necessary to consider the power flow of an eventual fusion reactor.
Here, we consider a steady-state reactor, so that the initial investment of power during the startup process contributes negligibly to the overall efficiency.
Such a power flow is shown in Fig.~\ref{fig:reactorPowerFlow}.
Electrical power $P_\text{in}$ consists both of power used to heat ($P_{H,e}$) and confine ($P_{C,e}$) the plasma.
With some conversion efficiency $\eta_H$, the electrical heating power is delivered to the plasma as heat $P_H = \eta_H P_{H,e} $.
As a result, the plasma produces some amount of fusion power $P_F$.
Meanwhile, power exits the plasma primarily through two possible mechanisms: bremsstrahlung radiation $P_B$, or thermal conduction loss $P_L$.
(We neglect for now other forms of radiation, such as electron cyclotron radiation, that depend on the magnetic field.
We also assume that bremsstrahlung is not reabsorbed, which is a safe assumption in the relatively low-density plasmas typical of  steady-state reactors.)
In steady state:
\begin{align}
	P_H + P_F = P_B + P_L. \label{eq:PBalanceReactionAbstract}
\end{align}
The relative balance between these terms is determined by the physics within the reactor.
Finally, the power that exits the plasma is converted back to into electrical power, with in general different efficiencies $\eta_B$ and $\eta_L$ for bremsstrahlung and thermal conduction loss respectively, resulting in a final output electrical power $P_\text{out} = \eta_B P_B + \eta_L P_L$.
Economical fusion energy requires that $P_\text{out}$ exceed $P_\text{in}$, preferably by a large margin.

The power flow here closely resembles that used in Wurzel and Hsu's recent analysis of progress towards fusion energy \cite{WurzelHsu2022Progress}.
There are three main differences here.
First, we have simplified the analysis of the heating energy by considering only a single conversion efficiency.
Second, we have explicitly separated out the electrical energy required for confinement.
Third, we have divided the output power into two streams with different electrical conversion efficiencies.
This last change reflects the fact that the aneutronic p-B11 reaction produces charged products, allowing for direct conversion of energy from lost particles, which has the potential to be much more efficient than the thermal processes likely required for conversion of bremsstrahlung energy.
Thus, keeping track of how power leaves the plasma is important to the overall energetic analysis.

The power leaving the reaction due to lost particles (i.e. thermal conduction) is generally written in terms of the confined kinetic energy density $U_K$ and the energy confinement time $\tau_E$:
\begin{align}
	\tau_E \equiv \frac{U_K}{P_L}. \label{eq:tauEDef}
\end{align}
Note that the power used to calculate this confinement time does not include the bremsstrahlung radiation $P_B$.
This formulation is convenient, as it generally leads to a requirement on the (temperature-dependent) product of density and confinement time $n \tau_E$, which is a useful fundamental target for fusion technology.
Achieving $P_\text{out} > P_\text{in}$ with a physically realizable $n \tau_E$ is the fundamental challenge of fusion energy science.

To measure the progress towards fusion, one generally looks at the $Q$ factor. 
There are several relevant $Q$ factors on the road towards economical fusion energy.
The ultimate goal is for a power plant to produce net power on the grid, determined by condition on the engineering $Q_\text{eng}$:
\begin{align}
	Q_\text{eng} &\equiv \frac{P_\text{out} - P_\text{in}}{P_\text{in}} > 0.
\end{align}
The higher $Q_\text{eng}$, the greater the ratio of power applied to the grid to recirculating power in the reactor.

Since we are looking at fundamental limits of the fusion efficiency, we will here consider a modified version of this metric, where we neglect the power used for confinement, i.e. assume $P_{C,e} = 0$.
We denote this modified $Q$ as $Q_\text{eng}^*$.
Then:
\begin{align}
	Q_\text{eng}^* = \bar{\eta} \lp Q_\text{fuel} + 1 \rp - 1, \label{eq:QEngStar}
\end{align}
where we have defined a quality factor associated with the fuel:
\begin{align}
	Q_\text{fuel} &\equiv \frac{P_F}{P_H}, \label{eq:QfuelDef}
\end{align}
and the average power recycling efficiency:
\begin{align}
	\bar{\eta} &\equiv \eta_H \lp \eta_L \frac{P_L}{P_L + P_B} + \eta_B \frac{P_B}{P_L + P_B} \rp < 1. \label{eq:avgRecyclingEfficiency}
\end{align}

High $Q_\text{fuel}$ is not a strictly necessary condition for net electricity production, if there is high recycling efficiency in the plasma.
Inverting Eq.~(\ref{eq:QEngStar}) and demanding $Q_\text{eng}^* > 0$ shows that net electricity production only requires:
\begin{align}
	Q_\text{fuel} > \frac{1}{\bar{\eta}} - 1, \label{eq:QfuelNetPower}
\end{align}
which can be small if the recycling efficiency is large, as can be the case with efficient direct conversion.
Nevertheless, achieving \emph{large} values of $Q^*_\text{eng}$ generally requires achieving even larger values of $Q_\text{fuel}$, making $Q_\text{fuel}$ a useful physics-based metric for the plasma performance.

\subsection{High-Performance Plasmas} \label{sec:highPerformanceDef}

If we want to focus on very high-performing plasmas, then, our goal is ultimately to obtain $Q_\text{fuel} \rightarrow \infty$.
This limit represents the state where the fusion reaction sustains itself without the need for external heating, known as burning plasma.

To look at what is necessary to achieve burning plasma, we use Eqs.~(\ref{eq:PBalanceReactionAbstract}), (\ref{eq:tauEDef}), and (\ref{eq:QfuelDef}) to rewrite  $Q_\text{fuel}$ as:
\begin{align}
	Q_\text{fuel} = \frac{P_F}{P_B + U_K/\tau_E - P_F}. \label{eq:Qfuel2}
\end{align}
Here, $P_F$, $P_B$, and $U_K$ are all determined immediately by the plasma parameters (the densities $n_s$ and temperatures $T_s$ of the species present), while $\tau_E$ depends on the details of the reactor design.
However, as a general rule, greater $n_i \tau_E$ is harder to achieve. 

Thus, as a general performance metric, we define $\tau_E^*$, the minimum value of $\tau_E$ (at a fixed $n_i$) that is required to achieve $Q_\text{fuel} \rightarrow \infty$.
From Eq.~(\ref{eq:Qfuel2}), this is given by:
\begin{align}
	\tau_E^* \equiv \tau_E \bigr|_{Q_\text{fuel} \rightarrow \infty} = \frac{U_K}{P_F - P_B}. \label{eq:tauEStarDef}
\end{align}

\subsection{The Dual Challenges of Thermonuclear p-B11 Fusion}

The quantity $\tau_E^*$ succinctly captures two main challenges that make proton-Boron 11 thermonuclear fusion--i.e. fusion with all species approximately Maxwellian--comparatively difficult.

First, we see from Eq.~(\ref{eq:tauEStarDef}) that $Q_\text{fuel} \rightarrow \infty$ requires the fusion power to exceed the bremsstrahlung power.
This has historically been a problem for p-B11 fusion, which requires large ion temperatures ($\sim$ 300 keV), and thus produces substantial bremsstrahlung, leading some to conclude that thermonuclear p-B11 fusion was infeasible \cite{Rider1995,Rider1995a}.
However, recent studies have indicated that the p-B11 cross section, particularly at high energies, is larger than previously thought \cite{Sikora2016CrossSection}.
A full energetic analysis by Putnvinski \emph{et al.}, considering collisional energy transfer between the various plasma species, revealed that these new cross sections opened up a small window where the fusion power could slightly exceed the bremsstrahlung power, making burning plasma theoretically achievable \cite{Putvinski2019}.
However, the margin by which $P_F$ exceeds $P_B$ is only a few percent at the most optimal parameters, making the prospect of burning plasma very difficult to envision with that energy balance.

Second, even in the absence of bremsstrahlung, the small cross section and high temperatures required for the reaction put a stringent limit on the confinement time.
To see this, note that the typical temperature of the reactants is around 300 keV (with around 150 keV for the electrons), while the typical fusion power per density product is:
\begin{align}
	U_\text{fus} \llangle \sigma_{pB} v \rrangle  = 4 \times 10^{-9} \text{eV cm$^{3}$ / s}.
\end{align}
Thus, even if bremsstrahlung were somehow suppressed, the maximum allowable $\tau_E^*$ at the optimal density $n_B / n_i = 0.15$ where $n_i = n_p+n_B$, is:
\begin{align}
	\tau_E^* = \frac{3}{2}\frac{n_i T_i + (n_B Z_B + n_p) T_e }{n_p n_B U_\text{fus} \llangle \sigma_{pB} v \rrangle }  \sim (16 \text{ s}) \lp \frac{10^{14} \text{ cm$^{-3}$}}{n_i} \rp.
\end{align}
Thus, at typical ITER densities, even in the absence of bremsstrahlung, the required energy confinement time is on the order of 16 seconds.
Given the results of Putvinski \emph{et al.}\cite{Putvinski2019}, the presence of bremsstrahlung makes this requirement $\sim 34$ times more stringent, i.e. $\tau_E \sim 540$ seconds.

\begin{figure*}[t]
	\centering
	\includegraphics[width=\linewidth]{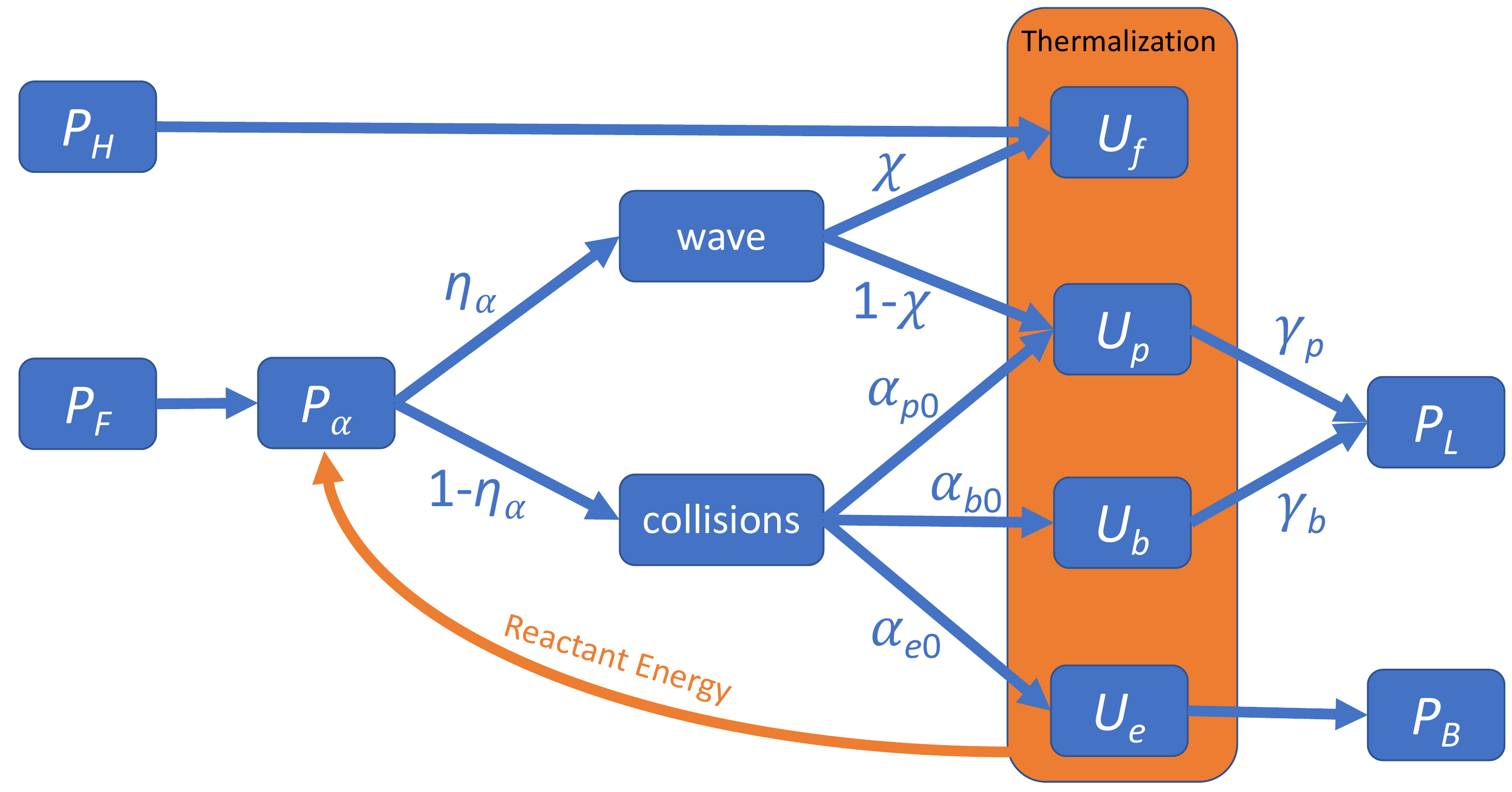}
	\caption{Simplified internal power flow model, relating the input heating power $P_H$ and fusion power $P_F$ to the output thermal conduction loss power $P_L$ and bremsstrahlung power $P_B$.
		Fusion energy $P_F$ combines with the power lost from the kinetic energy of the reactants and ends up in the alpha particles.
		The resulting alpha particle power $P_\alpha$ can be transferred into a wave with some efficiency $\eta_\alpha$, with a fraction $\chi$ of the wave power ending up in fast protons, and the remainder ending up in thermal protons.
		The remaining alpha particle power is collisionally transferred to thermal protons, boron, and electrons, in the fractions $\alpha_{p0}$, $\alpha_{b0}$, and $\alpha_{e0}$, respectively.
		In addition, collisional energy transfer and thermalization occur between all species.
		Energy is lost through thermal conduction $P_L$ from the proton and the boron populations in the fractions $\gamma_p$ and $\gamma_b$.
		Energy is lost from the electron population only through bremsstrahlung $P_B$.
	}
	\label{fig:internalPowerFlow}
\end{figure*}

\section{Internal Power Balance}\label{sec:powerBalanceModel}

The stringent requirements for thermonuclear p-B11 fusion encourage a consideration of nonthermal plasmas.
To examine the potential advantage such plasmas provide, we explore a power balance model similar to Putvinski \emph{et al.}\cite{Putvinski2019}, incorporating collisional temperature equilibration between species, fusion power production, and collisional transfer of alpha particle energy to the various thermal species.
However, to this balance of thermal protons $p$, boron $b$, and electrons $e$, we add a beam of monoenergetic fast protons $f$.
These fast protons can be maintained either by external energy input, or by using alpha channeling to transfer alpha power directly to the fast protons.
The power balance model thus takes the form:
\begin{align}
	\frac{dU_f}{dt} &= - K_{f p} E_f  - K_{f b} E_f - K_{f e} E_f \notag\\ 
	&\hspace{0.25in}- K_{F,f} E_f + \alpha_f P_\alpha + P_H \label{eq:dEfdt}\\
	\frac{dU_p}{dt} &= K_{f p} E_f + K_{pb} (T_b - T_p) + K_{p e} (T_e - T_p) \notag\\ 
	&\hspace{0.25in} - \frac{3}{2} K_{F,p} T_p + \alpha_p P_\alpha -\gamma_p P_L%\\
\end{align}
\begin{align}
	\frac{dU_b}{dt} &= K_{f b} E_f + K_{pb} (T_p - T_b) + K_{b e} (T_e - T_b)  \notag\\ 
	&\hspace{0.25in}-\frac{3}{2} (K_{F,f}+K_{F,p}) T_b + \alpha_b P_\alpha - \gamma_b P_L\\
	\frac{dU_e}{dt} &= K_{f e} E_f + K_{pe} (T_p - T_e) + K_{b e} (T_b - T_e) \notag\\ 
		&\hspace{0.25in}- P_B + \alpha_e P_\alpha. \label{eq:dTedt}
\end{align}
Here, we recognize the heating power $P_H$, thermal conduction loss power $P_L$, and bremsstrahlung power $P_B$.
This last can be approximated as\cite{Heitler2012Radiation,Putvinski2019}:
\begin{align}
	P_B &\approx 7.56 \times 10^{-11} n_e^2 x^{1/2} \bigl[ Z_\text{eff} \lp 1 + 1.78x^{1.34} \rp \notag\\
	& \hspace{0.2in} + 2.12 x \lp 1 + 1.1 x - 1.25 x^{2.5} \rp \bigr] \text{ eV cm$^{3}$/s,} \label{eq:PbrFormula}
\end{align}
where $x = T_e / E_\text{rest}$, $Z_\text{eff} = \sum_i n_i Z_i^2 / \sum_i n_i Z_i$, and $E_\text{rest} = 5.11\times 10^5$ eV is the electron rest energy.

In Eqs.~(\ref{eq:dEfdt}-\ref{eq:dTedt}), we have also defined many new variables.

First, $K_{ss'}$ for $s,s' \in \{f,p,b,e\}$ represents the energy transfer rate between species $s$ and $s'$. These rates are standard, but for completeness are described in Appendix~\ref{app:equilibrationRates}.

Second, $K_{F,f}$ and $K_{F,p}$ represent the fusion rate from fast and thermal protons respectively.
These rates include a kinetic enhancement factor to agree with Putvinski \emph{et. al.}\cite{Putvinski2019} in the appropriate limits, and are described in Appendix~\ref{app:fusionRates}.

Third, in agreement with Putvinski, we assume that power is lost through thermal conduction only from the ions.
Since we track both boron and proton temperatures separately, we must choose how to partition this loss, which is encoded in the parameters $\gamma_p$ and $\gamma_b$.
We assume that thermal losses in each thermal ion species occur proportionally to that species' pressure:
\begin{align}
	\gamma_i \equiv \frac{n_i T_i}{\sum_{j} n_{j} T_{j}}, \quad i,j\in \{p,b\}.
\end{align}

Fourth, we have defined a new power $P_\alpha$, given by:
\begin{align}
	P_\alpha &\equiv \mathcal{E}_F (K_{F,f} + K_{F,p}) + K_{F,f} \lp E_f + \frac{3}{2} T_b \rp\notag\\
	& \hspace{1.13in} + \frac{3}{2} K_{F,p} (T_p + T_b),
\end{align}
which represents the typical power flowing through the alpha particles, determined by the sum of the fusion energy per reaction ($\mathcal{E}_F = 8.7$ MeV) and the initial energy of the fusing particles.
This power is distinct from the fusion power, which is given by just the first term:
\begin{align}
	P_F = \mathcal{E}_F (K_{F,f} + K_{F,p}).
\end{align}
Note that the contribution to $P_\alpha$ from the thermal species is an approximation, since in general the average reactant energy will not necessarily be the same as the thermal average energy.

Finally, we have defined the fraction of alpha particle power $\alpha_s$ that goes into each species.
To model alpha channeling, we assume that some determined fraction $\eta_\alpha$ can be redirected by waves into the protons.
Of this wave energy, we assume that a fraction $\chi$ ends up in the fast protons $f$, with $(1-\chi)$ going to the thermal protons $p$.
Note that this model assumes that the wave energy is fully damped in the plasma.
For the remaining alpha particle energy, we assume that it is partitioned between the remaining species according to rate at which alpha particles transfer energy collisionally to that species, as determined by the parameter
\begin{align}
	\alpha_{s0} \equiv \llangle \frac{K_{\alpha s}}{K_{\alpha p} + K_{\alpha b} + K_{\alpha e}} \rrangle, \quad s \in \{p,b,e\}.
\end{align}
Here, the average is performed over the hot alpha particle distribution, as described in Appendix~\ref{app:alphaCollisions}.
Thus, the total fraction of alpha particle power going to each species, including both alpha channeling and collisional effects, is given by:
\begin{align}
	\alpha_f &= \eta_\alpha \chi \label{eq:alphaF}\\
	\alpha_p &= (1-\eta_\alpha) \alpha_{f0} + \eta_\alpha(1-\chi)\\
	\alpha_b &= (1-\eta_\alpha) \alpha_{b0}\\
	\alpha_e &= (1-\eta_\alpha) \alpha_{e0}. \label{eq:alphaE}
\end{align}
The overall power flow represented by Eqs.~(\ref{eq:dEfdt}-\ref{eq:dTedt}) and (\ref{eq:alphaF}-\ref{eq:alphaE}), incorporating both collisions and alpha channeling, is schematically represented in Fig.~\ref{fig:internalPowerFlow}.

%To calculate $K_{\alpha s}$, we use the typical alpha particle energy:
%\begin{align}
%	E_\alpha = \frac{1}{3} \left[\mathcal{E}_F + \frac{K_{F,f} E_f}{K_{F,p} + K_{F,f}}  + \frac{3}{2}\frac{K_{F,p} T_p }{K_{F,p} + K_{F,f}} + \frac{3}{2} T_b \right].
%\end{align}
%This is an approximation, since in any practical system there will be a range of alpha particle energies, and so the alpha particle energy is a random variable given by a distribution that depends on the fusion sources, collisional kinetics, and the details of the channeling and confinement processes.

\subsection{Verification of Power Balance Model} \label{sec:putvinskiVerification}

\begin{figure}[t]
	\centering
	\includegraphics[width=\linewidth]{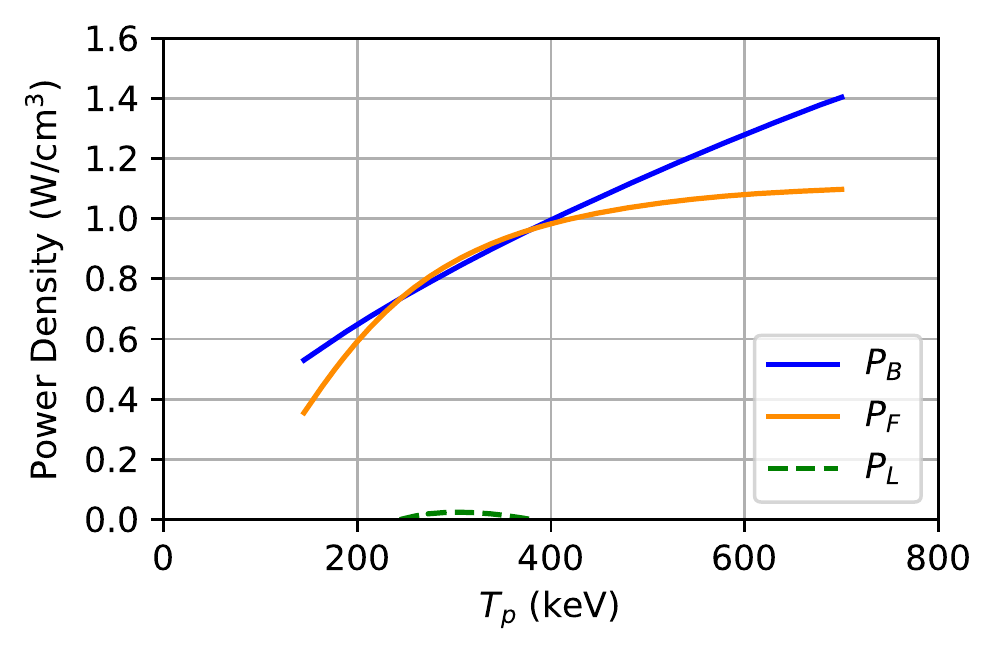}
	\caption{Power balance results for a mix of 15\% Boron, 85\% thermal protons, the same case considered by Putvinski\cite{Putvinski2019}. Despite the coarse approximations to the kinetic physics, the agreement is quite good. Fusion power exceeds bremsstrahlung power by a maximum of around 3\% around 300 keV.}
	\label{fig:putvinskiVerification}
\end{figure}

To check our power balance model (with its coarse approximations of the kinetic physics), we first check whether it recovers the basic results of Putvinski's power balance analysis.
To this end, we set the boron and proton fractions $n_B = 0.15 n_i$, $n_p = 0.85 n_i$, with $\eta_\alpha = n_f = 0$, and solved for the power balance for a range of proton temperatures $T_p$.
The results are shown in Fig.~\ref{fig:putvinskiVerification}, which can be compared with Putvinski's Fig.~4.
The agreement is quite good: in each case, the fusion power exceeds the bremsstrahlung power by at most $\sim 2.8\%$, at around $300$ keV.

At this point, it is useful to note that although the curves in Fig.~\ref{fig:putvinskiVerification} represent steady-state solutions to the system of differential equations in Eqs.~(\ref{eq:dEfdt}-\ref{eq:dTedt}), not all of these solutions are stable.
For temperatures above the peak value of $P_L$ at 300 keV, a decrease in thermal conduction losses will heat the plasma, thus decreasing the gap between $P_F$ and $P_B$ until it matches the new level of $P_L$, thus achieving a new steady state. 
Conversely, at temperatures below 300 keV, a decrease in $P_L$ losses will still heat the plasma, driving it into a region where there is an even larger gap between $P_F$ and $P_B$, heating the plasma even more rapidly.
Thus, in this case, the temperature will increase until it reaches the stable operating point for the given value of $P_L$, at a temperature above 300 keV.
This general stability property, where solutions are stable at temperatures above the maximum value of $(P_F-P_B)$ and unstable below it, will hold for the solutions throughout the paper.
As a consequence, the operating temperature of an igniting reactor that exceeds the minimal requirements for ignition will be somewhat hotter than the optimal temperature that maximizes the allowable thermal conduction losses $P_L$.
It is also important to note that an economical reactor would not be run in an ignited mode; in order to exercise control over the plasma, it would always be advantageous to maintain the plasma just below ignition, which assures large multiplication to any auxiliary power.

\section{Optimizing the Power Balance with Alpha Channeling} \label{sec:tauEOptimization}

Having described the internal and reactor-wide power balance, we now move on to a description of the optimization of the reaction given different assumptions for the alpha channeling, both with and without channeling into fast ions.
This allows us to determine the potential upside of alpha channeling in improving the feasibility of the p-B11 reaction.

\subsection{Free and Determined Parameters}

For the power balance in Eqs.~(\ref{eq:dEfdt}-\ref{eq:dTedt}) we have 12 parameters: $n_s$ for $s \in \{f,p,b,e\}$, $E_f$, $T_s$ for $s \in \{p,b,e\}$, $\eta_\alpha$, $\chi$, $P_H$, and $P_L$.
In steady state, we must have $dU_s / dt = 0$, defining four constraint equations.
Thus, the steady-state solutions lie on an 8-dimensional manifold in a 12-dimensional space.
However, we generally assume quasineutrality:
\begin{align}
	n_e = n_f + n_p + Z_b n_b,
\end{align}
which adds a constraint, reducing the solution manifold to 7 dimensions.

Our eventual goal in developing a reactor is to optimize $Q_\text{eng}^*$ on this manifold.
Generally, this optimization is done at a set fuel ion density, since all terms in the optimization scale as $n_i^2$, except for logarithmic scaling in the collision terms.
We also expect to optimize separately for each possible alpha channeling efficiency $\eta_\alpha$, and each possible split of channeling energy between fast and thermal protons $\chi$.
These considerations add three constraints, reducing the solution manifold on which the optimization must be performed to 4 dimensions.

Unfortunately, the optimization of $Q_\text{eng}^*$ depends on the three engineering parameters $\eta_H$, $\eta_B$, and $\eta_L$.
Thus, a different optimization must be performed for each combination of the various energy conversion efficiencies, dramatically expanding the problem space and reducing the interpretability of the results.

Thus, for simplicity, generality, and comparison to earlier work \cite{Putvinski2019}, we focus instead on the space of high-performance, burning plasma operation described in Section~\ref{sec:highPerformanceDef}.
Therefore, we aim to minimize the required energy confinement time $\tau_E^* = U_K / P_L$ for $Q_\text{fuel} \rightarrow \infty$, at a set value of the ion density $n_i$, channeling efficiency $\eta_\alpha$, and channeling fraction to fast protons $\chi$.
In addition to eliminating the dependence of the result on the various engineering $\eta$'s, this optimization provides another constraint, since $Q_\text{eng}^* \rightarrow \infty$ requires that $P_H \rightarrow 0$.
Thus, the eventual optimization occurs on a 3-dimensional manifold in the 12-dimensional parameter space.

For each value of $\eta_\alpha$ and $\chi$, the optimization of $\tau_E^*$ is performed numerically over the variables $n_p$, $n_b$, $E_f$, and $T_p$, with the remaining variables ($n_e$, $T_b$, $T_e$, and $P_L$) solved for using the constraints given by quasineutrality and the power balance.
The optimization is performed using the Sequential Least Squared Programming (SLSQP) algorithm implemented as an option in to scipy.minimize. 
This method allows for both inequality constraints, required to keep the temperatures and densities positive, as well as equality constraints, necessary for enforcing $n_i \equiv n_f + n_b + n_p = 10^{14}$ cm$^{-3}$.

The power balance and optimization here share many similar features to those in work by Hay\cite{Hay2015Ignition}.
However, that paper only incorporated power loss due to bremsstrahlung, ignoring possible thermal conduction losses.
Thus, from the perspective of reactor design, the results were somewhat overconstrained, representing the boundary of the ignition region with $P_L = 0$, rather than the region where $P_L > 0$ at which ignition can occur even with thermal conduction losses.

%\begin{figure*}[t]
%	\centering
%	\includegraphics[width=0.8\linewidth]{putvinskiVsChanneling.png}
%	\caption{adsf}
%	\label{fig:putvinskiVsChanneling}
%\end{figure*}

\subsection{Improvement in Confinement Times for Ignition} 

We now turn to the results of the optimization.
For simplicity in the following discussion, we primarily consider the cases of $\chi = 0$ (alpha channeling only to thermal protons), and $\chi = 1$ (alpha channeling only to fast protons), since the intermediate cases basically interpolate between these extremes.

Our first finding is that, interestingly, in the absence of channeling, $\tau_E^*$ is not optimized at the classic value of 15\% boron, 85\% protons. 
Instead, it is optimized at 13\% boron, 87\% proton, which reduces $\tau_E^*$ from around 550 seconds to around 460 seconds.

\begin{figure}[t]
	\centering
	\includegraphics[width=\linewidth]{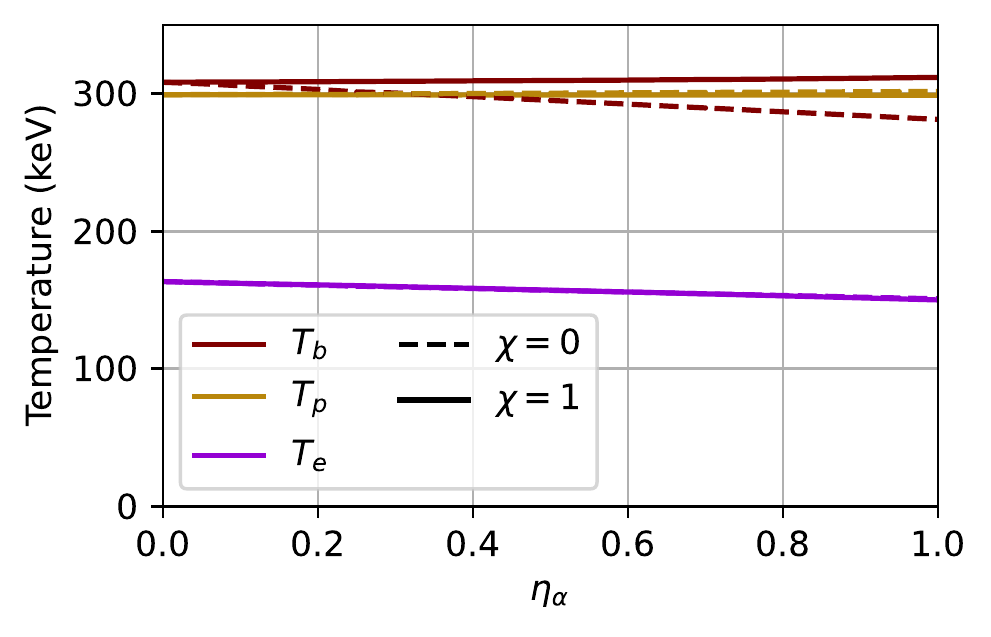}
	\caption{Temperature of different species with changing fractions of fusion power $\eta_\alpha$ channeled to fast particles.
		The protons heat up, while the boron and electrons cool.
	However, the boron cooling is more pronounced with $\chi = 0$.}
	\label{fig:TVsEta}
\end{figure}

As $\eta_\alpha$ increases, regardless of $\chi$, the optimal thermal proton temperature remains unchanged at 300 keV, while the optimal electron temperature drops from its initial value of 163 keV to 150 keV, reflecting the fact that less alpha particle energy is directly transferred to electrons (Fig.~\ref{fig:TVsEta}).
However, the corresponding boron temperature depends on $\chi$.
When $\chi = 1$, the boron temperature remains fairly flat, reflecting the fact that the fast protons collisionally heat the boron in much the same manner as the alpha particles. 
In contrast, when $\chi = 0$, the optimal boron temperature drops by around 30 keV to 281 keV, below the proton temperature.

\begin{figure}[h]
	\centering
	\includegraphics[width=\linewidth]{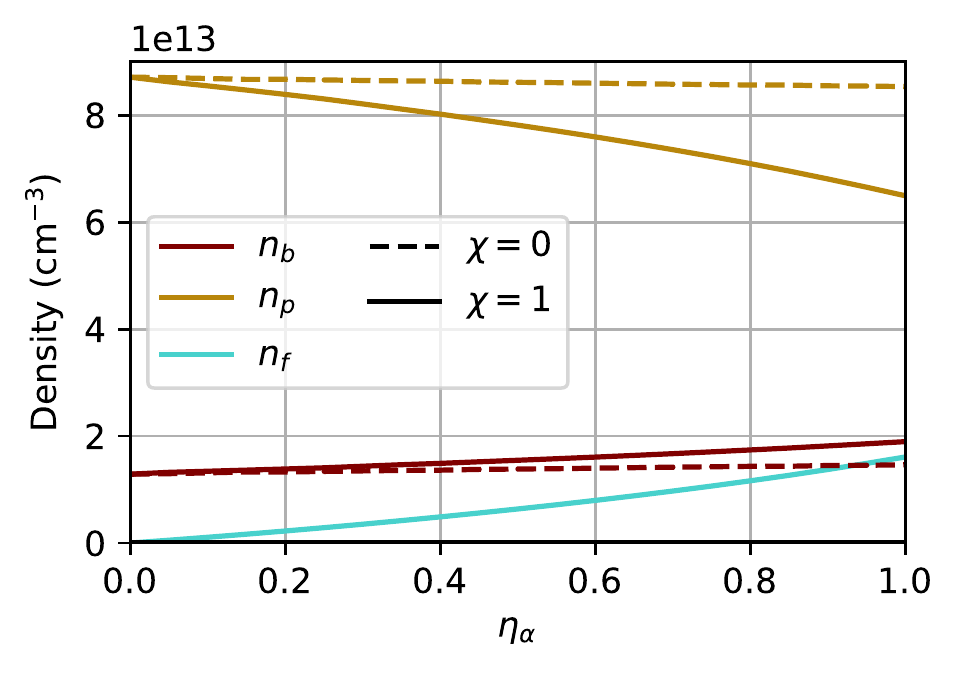}
	\caption{Optimal density of different species with changing fractions of fusion power $\eta_\alpha$ channeled to thermal protons ($\chi = 0$) or fast particles ($\chi = 1$).
	For $\chi = 0$, the initial 13\% boron 87\% proton mix changes slightly to 15\% boron, 85\% protons.
	Meanwhile, for $\chi = 1$, the optimal boron concentration becomes much larger, rising to 19\% boron at $\eta_\alpha = 1$, with 65\% thermal protons and 16\% fast protons.}
	\label{fig:nVsEta}
\end{figure}

\begin{figure}[h]
	\centering
	\includegraphics[width=\linewidth]{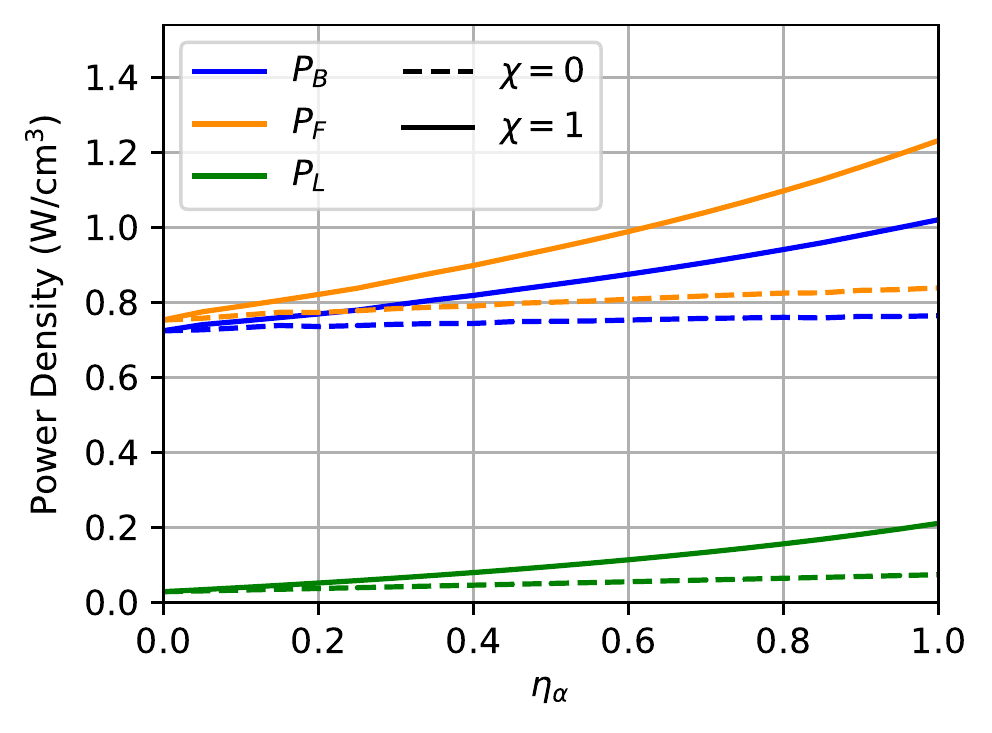}
	\caption{Change in optimal fusion and bremsstrahlung power with increased alpha channeling efficiency $\eta_\alpha$. 
		Results are shown for the case where power is channeled into fast protons ($\chi = 1$) vs. thermal protons ($\chi = 0$).
		In both cases, increased $\eta_f$ leads to a larger difference $P_L = P_F - P_B$ than in the case without alpha channeling; however, the difference grows much faster when channeling to fast protons.}
	\label{fig:PVsEta}
\end{figure}

\begin{figure}[t]
	\centering
	\includegraphics[width=\linewidth]{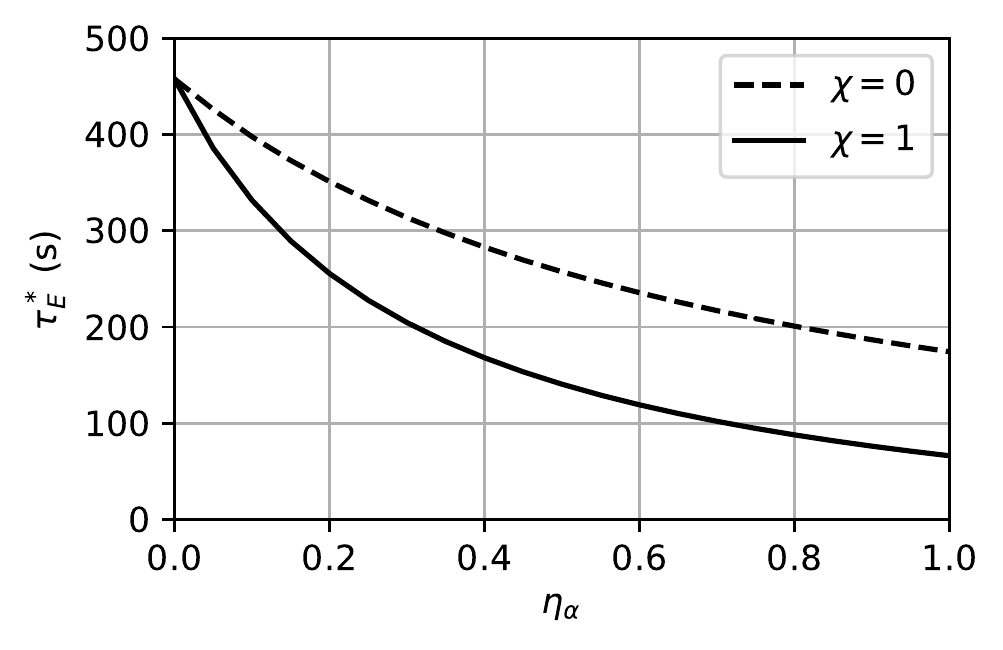}
	\caption{Change in energy confinement time required for ignition $\tau_E^*$ (at ion density $n_i = 10^{14}$ cm$^{-3}$) for different values of the alpha channeling fraction $\eta_\alpha$. 
	When channeling into thermal protons ($\chi = 0$), 50\% efficient alpha channeling results in a 1.8x reduction in $\tau_E^*$ to 250 seconds, while 100\% efficient channeling results in a 2.6x reduction to 170 seconds.
	Channeling into fast protons is even more effective: channeling 50\% of the alpha power into fast protons results in a 3.2x reduction in $\tau_E^*$ to 140 seconds, while channeling 100\% of the power results in a 6.8x reduction to 66 seconds.}
	\label{fig:tauEVsFf}
\end{figure}

\begin{figure*}[t]
	\centering
	\includegraphics[width=\linewidth]{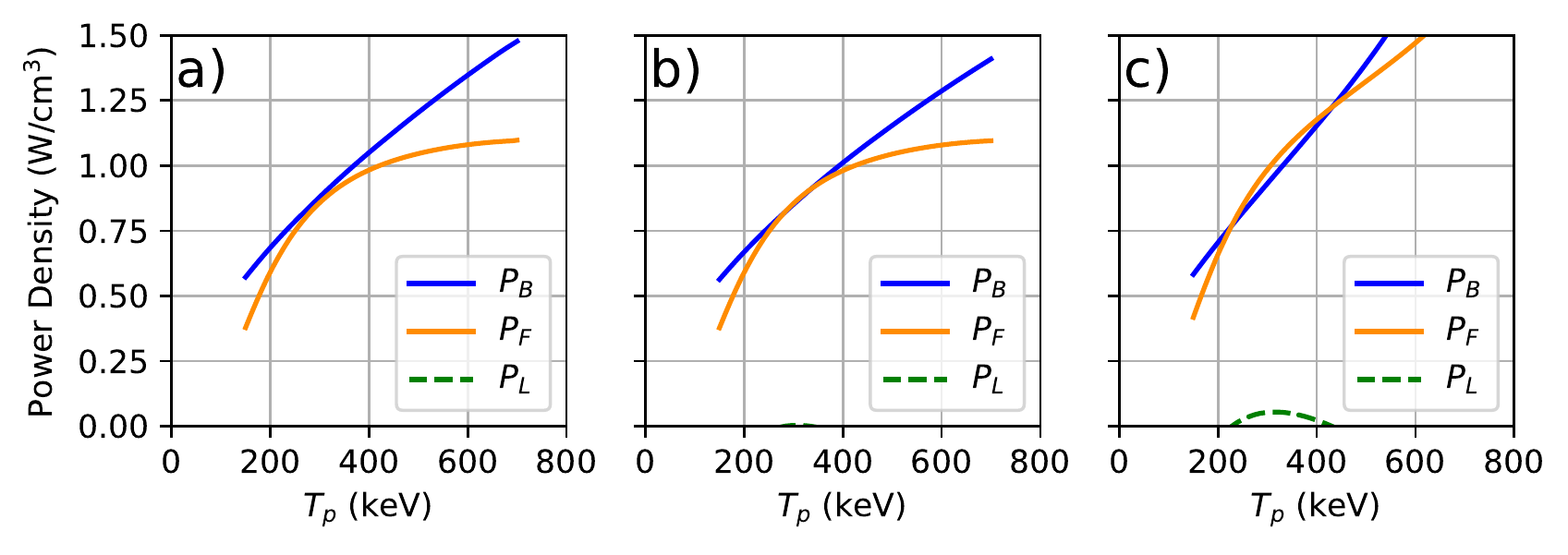}
	\caption{(a) Bremsstrahlung power $P_B$ and fusion power $P_F$ as a function of proton temperature for a base ion mixture of 15\% boron, 85\% thermal protons, with an additional 3\% of alpha particles added. The bremsstrahlung power is higher than the fusion power, making ignition impossible.
	(b) The same case, but with 50\% of the alpha power channeled into thermal ions.
	The bremsstrahlung power is reduced, while the fusion power increases, so that $P_F > P_B$ for a window around 350 keV.
	(c) The same case, but with 50\% of the alpha power channeled into an added population of 650 keV fast ions (with fast ion density determined by the power balance constraints---around 6.5\% $n_i$ at 300 keV).
	Both bremsstrahlung and fusion power increase, but below $T_p \sim 450$ keV, the bremsstrahlung power increases more slowly than the fusion power, so that $P_F > P_B$, making ignition possible.}
	\label{fig:alphaPoisoning}
\end{figure*}

In light of these changes in temperature, the optimal ion mix changes as well.
For the case without fast ions ($\chi = 0$), the Boron fraction increases from $13\%$ to $15\%$ (Fig.~\ref{fig:nVsEta}).
This leads to a rise in bremsstrahlung power, which is somewhat mitigated by the drop in electron temperature, leading to a net 6\% increase in bremsstrahlung power (Fig.~\ref{fig:PVsEta}).
However, the $\sim 13\%$ increase in boron fraction leads to an overall 11\% increase in the fusion power, increasing $P_L \equiv P_F - P_B$ from 4\% of $P_F$ to 9\% of $P_F$.

For the case of $\chi = 1$, the optimal ion mix changes more dramatically.
Of course, as more power is channeled into the fast protons, the fast proton density increases, rising to 16\% at $\eta_\alpha = 1$.
At the same time, the optimal boron fraction also rises from its initial value of 13\% to 19\% (Fig.~\ref{fig:nVsEta}).
This increase in boron density causes the bremsstrahlung power to rise $41\%$ despite the drop in electron temperature; however, at the same time, the fusion power rises by a massive 64\%, leading to a large net increase in $P_L$ from 4\% of $P_F$ to  17\% of $P_F$ (Fig.~\ref{fig:PVsEta}).

These increases in $P_L$ lead to an order-of-magnitude improvement in the required energy confinement time for ignition $\tau_E^*$.
Even without channeling into fast ions ($\chi = 0$), $\tau_E^*$ falls rapidly with increasing $\eta_\alpha$, from 460 seconds without alpha channeling to 170 seconds at $\eta_\alpha = 100\%$.
Channeling into fast protons ($\chi = 1$) improves these results even further, bringing $\tau_E^*$ down to 66 seconds at $\eta_\alpha = 1$ (Fig~\ref{fig:tauEVsFf}).
For general $\eta_\alpha$ and $\chi$, the improvements to $\tau_E$ can be fit by a simple analytic function to within 7\%:
\begin{align}
	\tau_E^* &= \frac{\tau_{E0}^*}{1+\eta_\alpha^{1.24}(1.72+3.85\chi^{1.15})}.
\end{align}
Here, $\tau_{E0}^* = 458$ seconds is the value of $\tau_E^*$ at $\eta_\alpha = 0$. 

\subsection{Alpha Particle Loss in Steady State}

For a steady-state reactor, the alpha particle loss rate is related to the fusion power by:
\begin{align}
	\dot{n}_\alpha &= 3 P_F / \mathcal{E}_F.
\end{align}
If alpha particles are lost at the ambient ion temperature, then the energy lost through this mechanism is:
\begin{align}
	P_{L\alpha} &= \frac{3}{2} T_\alpha \dot{n}_\alpha = \frac{9}{2} \frac{T_\alpha}{\mathcal{E}_F} P_F \approx 15.5\% P_F.
\end{align}

This result represents a huge problem for thermonuclear fusion, which can tolerate only $P_L/P_F = 3\% $.
It also represents a substantial challenge for alpha-channeling-driven reactions, since 16\% losses use up almost the entire allowable thermal conduction losses even for 100\% efficient alpha channeling into the fast protons (with $P_L/P_F = 17\%)$.

However, it is important to remember that alpha particles do not necessarily have to be released at the thermal temperature. 
There are a variety of mechanisms which might be used to preferentially release the particles at lower energy.
For instance, in simple magnetic mirrors, the low-energy ions tend to exit much more quickly than those at high energy, since they scatter into the loss cone faster.

In fact, alpha channeling provides a natural mechanism to cool the alpha particles below the thermal temperature, while transferring the energy into fuel  which are driven into the plasma core. 
This works because the spatial position of a particle is coupled to its momentum, which is in turn proportional to energy absorbed by the wave.
Thus, in steady state, particles which diffuse outward and lose energy (alphas) are counteracted by those which diffuse inward and gain energy (fuel ions).

The more generous margins values of $P_L / P_F$ afforded by alpha channeling also reduce the requirements on the energy of the lost alpha particles.
For purely thermonuclear fusion, with $P_L/P_F = 3\% $, alpha particles must be lost with an effective temperature of less than 20\% of the bulk ion temperature; for 100\% efficient alpha channeling into thermal protons with $P_L/P_F = 9\% $, this figure becomes 60\%, while for fast protons it can be over 100\%.
Thus, the advantages of alpha channeling become even more dramatic when the requirements on particle fluxes in steady state are appreciated.

\section{Robustness to Ash Poisoning} \label{sec:ash}

In the power balance above, as in Putvinski \cite{Putvinski2019}, it was assumed that the ash population was negligible; however, in a realistic reaction, some amount of ash will always be present.
This has the potential to quickly close the ignition window for thermonuclear fusion, since it is already so marginal.
Even as small an ash concentration as $2\%$ $n_i$ is enough to make $P_F < P_B$, necessitating external heating power input to keep the reaction going even with perfect energy confinement.

Because transfer of alpha power to fast ions increases fusion power relative to bremsstrahlung power, it also increases the robustness of the ignition conditions to poisoning by alpha particle ash (or other impurities).
To see this, we can add a population of alpha particles to the power balance, which only contribute to the bremsstrahlung energy.
We can then compare the fusion and bremsstrahlung yields of a 15\% boron-85\% thermal proton base mix, with some contamination by fusion ash, to the case where some portion of the alpha particles goes to support an additional population of fast protons.
Note that such an analysis is not optimized, nor does it have constant $n_i$ across all cases; nevertheless, it gives us a clear picture of the effect of alpha channeling.

In Fig.~\ref{fig:alphaPoisoning}, we show the result of this analysis for a 3\% $n_i$ population of alpha particles, (a) in the absence of alpha channeling, (b) with 50\% efficient ($\eta_\alpha = 0.5$) alpha channeling into thermal protons ($\chi = 0$), and (c) with 50\% efficient ($\eta_\alpha = 0.5$) alpha channeling into fast protons ($\chi = 1$).
While the bremsstrahlung power definitely exceeds the fusion power in the case without channeling (Fig.~\ref{fig:alphaPoisoning}a), in both cases with channeling (Fig.~\ref{fig:alphaPoisoning}b-c), fusion power exceeds bremsstrahlung power; in fact, in the case of channeling by fast protons, by a larger margin than in the case with no ash poisoning and no alpha channeling (Fig.~\ref{fig:putvinskiVerification}).
Thus, in addition to dramatically reducing the required confinement energy time for ignition, the ability to channel energy directly from alpha particles to fast protons makes the ignition condition far more robust to poisoning by impurities and ash.

\section{Generalized Beam-Target Fusion} \label{sec:beamTarget}

One might wonder why we bother with the thermal proton population at all, if the high-energy protons are so much more reactive.
In other words, why not perform a form of generalized beam-target fusion, where alpha energy is channeled back into the fast protons, and the other species are kept cold to limit bremsstrahlung.

The reason that this paradigm does not work, at least to achieve ignition, is that the fast protons cannot provide enough fusion energy to support themselves against the slowing down energy of the reactor. 
The peak power density of the p-B11 reaction, for 650 keV protons impacting cold boron, is\cite{Sikora2016CrossSection}:
\begin{align}
	P_{F,\text{max}} &= v_p \sigma_{pb} n_p n_b \bigl|_{650 \text{keV}} \\
	&= 1.4 \times 10^{-8} n_p n_b \; \; \text{eV$\,$cm$^{3}$/s}, \label{eq:BeamPfmax}
%	&= 1.8 \times 10^{-9} n_i^2 \; \; \text{eV$\,$cm$^{3}$/s}, \label{eq:BeamPfmax}
\end{align}
%where $n_b$ and $n_p$ are in cm$^{-3}$.
where $n_i$ is in cm$^{-3}$.
%Importantly, this cross section falls off with increasing $T_b$, losing around $10^{-9}$ eV$\,$cm$^{3}$/s for every 30 keV in $T_b$. 

Meanwhile, the power transfer from the fast protons into the cold boron, with $E_f$ in eV and $m_p$ and $m_b$ in proton mass units, is:
\begin{align}
	K_{fb} E_f &\approx 1.8 \times 10^{-7} Z_p^2 Z_b^2 \frac{m_p^{1/2}}{m_b} \lambda_{pb} E_f^{-1/2} n_p n_b\\
	&\approx 1.2 \times 10^{-8} n_p n_b \; \; \text{eV$\,$cm$^{3}$/s}.
%	&\approx 1.5 \times 10^{-9} n_i^2 \; \; \text{eV$\,$cm$^{3}$/s}.
\end{align}
%Combined with Eq.~(\ref{eq:BeamPfmax}) and the scaling of the fusion power with increasing boron temperature, this implies that $T_b \lesssim 50$ keV, in order for fusion power to be able to compensate supply the power damped on boron by the fast protons.
Thus, an irreducible, large fraction of the fusion power used to support the beam is already used up in supporting the beam against collisions with the boron.

The beam does not only encounter boron, but also electrons.
The power transfer from the proton beam into the electrons is given by:
\begin{align}
	K_{fe} E_f &\approx 3.2 \times 10^{-9} Z_p^2 m_p^{-1} \lambda_{pe} T_e^{-3/2} E_f n_p (n_p+Z_b n_b)\\
	&\approx 3 \times 10^{-7} T_e^{-3/2} E_f n_p (n_b + n_p/Z_b)  \; \; \text{eV$\,$cm$^{3}$/s}.
%	&\approx 4.5 \times 10^{-7} T_e^{-3/2} E_f n_i^2  \; \; \text{eV$\,$cm$^{3}$/s}.
\end{align}
For the fast protons to be supported against collisional slowing, $K_{fe} E_f$ has to be smaller than $P_{F,\text{max}} - K_{fb} E_f$.
This pushes the electrons towards higher temperatures.
However, at the same time, the electrons must be heated against bremsstrahlung radiation, i.e. $K_{fe} E_f$ has to be larger than $P_B$.
As the electron temperature increases and $K_{fe} E_f$ decreases, $P_B$ increases, eventually overtaking it.
Regardless of the optimizations one tries with the mix of proton and boron, these multiple constraints make this self-sustaining beam fusion impossible at densities in the $10^{14}$ cm$^{-3}$ range, even when assuming 100\% alpha channeling efficiency.
Employing a hybrid scheme, with both thermal and fast proton populations, as in Section~\ref{sec:tauEOptimization}, serves to relax these harsh constraints, leveraging both the high cross section of the fast protons and the contribution of the thermal protons to open up a wider regime of ignition scenarios.
These issues are discussed at more length in a second publication \cite{kolmes2022waveSupported}, which uses a simple analytical model to clarify the conditions under which such a hybrid scheme is favorable.

Interestingly, the beam fusion requirements relax at high density, since the Coulomb logarithm decreases significantly.
For instance, for a 14\% boron / 86\% fast proton mix at $n_i = 10^{19}$ cm$^{-3}$, with the boron at 10 keV and the electrons at 140 keV, it is true both that $P_F > K_{fb} E_f + K_{fe} E_f$, and that $K_{fe} E_f > P_B$.
However, ignition in this case would still require extremely high alpha channeling conversion efficiency and confinement time, now occurring at extremely high densities, and so remains impractical.

\section{Recycling and Net Power Output} \label{sec:recycling}

The inability to ignite the plasma, either because it is fundamentally unachievable (as in the case of beam-target fusion) or because it requires too stringent confinement times (as might be the case for hybrid fast-ion thermonuclear fusion of Section~\ref{sec:tauEOptimization}) does not necessarily mean that net energy production, i.e. $Q_\text{eng}^* > 0$, is unachievable.
As demonstrated in Eq.~(\ref{eq:QfuelNetPower}), the fusion power output can be lower than the heating power, as long as the power recycling efficiency $\bar{\eta}$ is high.
These high recycling efficiencies are likely to be a particular strength of aneutronic fuels such as p-B11, which produce charged products that stay in the plasma, opening the door to high-efficiency direct conversion of their power to electric energy.

To see how direct conversion might make net energy production feasible even with $\tau_E \ll \tau_E^*$, consider a reactor where the heating power $P_H$ is chosen to offset the excess thermal conduction losses $(P_L - E_K/\tau_E^*)$.
If $P_H$ is put back into the same species that $P_L$ is lost from, then this leaves us with the same self-consistent equilibrium as for the case with $P_H = 0$.
For this equilibrium, taking $\eta_H = 1$ so that $\eta_B$ and $\eta_L$ now represent recycling efficiencies, $Q_\text{eng}^*$ is given from Eq.~\ref{eq:QEngStar} by:
\begin{align}
	Q_\text{eng}^* &= \frac{\eta_B P_B + \eta_L (E_K/\tau_E)}{E_K/\tau_E - E_K/\tau_E^*} - 1.
\end{align}
Thus, the condition $Q_\text{eng}^* > 0$ becomes:
\begin{align}
	\tau_E > \tau_E^* (1-\eta_L) \lp 1 + \frac{\eta_B P_B}{P_L^*} \rp^{-1}, \label{eq:tauERecycling}
\end{align}
where $P_L^* = E_K / \tau_E^*$. 
Thus, with thermal conduction loss efficiencies of 80\%-90\%, a reactor could sustain a 5-10x lower confinement time, even if none of the bremsstrahlung energy was recycled.
With recycling of the bremsstrahlung energy, this can likely be reduced by a further 2-3x.
Thus, the equilibrium that requires a 66 second confinement time for ignition could require as little as 2 seconds for net power production.

It should be noted that recycling even allows net power production at $\tau_E^*$ when $\tau_E^* = 0$, which can be seen from a rearrangement of Eq.~\ref{eq:tauERecycling}:
\begin{align}
	\tau_E > \frac{E_K}{\eta_B P_B} (1-\eta_L) \lp 1 + \frac{P_L^*}{\eta_B P_B} \rp^{-1}.
\end{align}
This formulation makes it clear that alpha channeling can significantly aid a reactor in achieving net power production when it can make the maximal loss power allowable for ignition $P_L^*$ comparable to the recycled heating power from bremsstrahlung radiation $\eta_B P_B$.

Of course, this discussion assumed that the same equilibrium was optimal for both ignition and high-$Q_\text{eng}^*$ operation with recycling.
As shown in Eq.~(\ref{eq:avgRecyclingEfficiency}), the average recycling efficiency $\bar{\eta}$ is a weighted ratio of the bremsstrahlung and charged particle electrical conversion energies.
Since direct conversion is likely to make energy from particle loss very efficient, the former is likely much smaller than the latter.
Thus, above a certain direct conversion efficiency, it could be better to quickly dump power out of the plasma through particle loss, to keep the bremsstrahlung low and increase the overall recycling efficiency.
Indeed, this is the approach taken by Volosov\cite{Volosov2006ACT,Volosov2011Problems}, which employs deliberately large loss cones in a mirror confinement setup, and then re-harvests the lost particle energy through an arrangement of concentric end electrodes.
Of course, this scheme requires more heating power to be applied to the plasma, and thus lower $Q_\text{fuel}$; but as long as the heating power can be provided by recycling the lost power, it still allows for net energy output.
Thus, the optimal operating point for high $Q_\text{eng}^*$ might be quite different from that optimal for ignition, and might be achieved at even lower $\tau_E$.

A detailed evaluation of the possible effects of alpha channeling on reactors with recycling is a large topic in and of itself, since there will be a different optimization (and different results) for every possible set of efficiencies $\eta_H$, $\eta_L$, and $\eta_B$, as well as for each different achievable energy confinement time $\tau_E$.
Nevertheless, it is important to note that power recycling has the potential to dramatically reduce the requirements for economical p-B11 fusion.

%Discuss ACT, Volosov, short discussion of recycling efficiency requirements.

\section{Additional Power Loss Mechanisms} \label{sec:addlPowerLoss}

Our discussion so far has focused on the fundamentals of the power balance that are fairly independent of confinement scheme, since all the terms scale roughly as $n^2$, and thus depend only on the various species' temperatures (or energies, for the fast protons).
Here, we discuss a couple other important terms in the power balance, which are more device-specific.

\subsection{Confinement Power}

In the original power flow in Fig.~\ref{fig:reactorPowerFlow}, power used to maintain the confinement represented an important component of the power flow, which was ignored in the subsequent analysis.
Although the power required to maintain all the systems involved in the confinement will be very device-dependent, it is useful to discuss briefly and generally the constraints that result from confinement considerations. 

Usually, the confinement will involve the use of magnetic fields.
The ratio of the plasma pressure to the magnetic pressure forms the plasma $\beta$:
\begin{align}
	\beta &\equiv \frac{P}{B^2 / 8\pi} = \frac{2}{3} \frac{U_K}{U_C},
\end{align}
where in the last line we have written the $\beta$ as a ratio of confined thermal energy density $U_K$ to field energy $U_C = B^2/8\pi$.
If the confinement system also involves electric fields and flows with significant energy, then $U_C$ can be generalized to $U_C = (E^2 + B^2)/8\pi + \sum_s n_s m_s u_s^2 /2$, where $u_s$ is the fluid velocity.

In general, the energy in these fields, just like the energy in the plasma, will decay, from processes such as resistivity in the coils, conductivity mechanisms, and viscosity, and must be supported by a constant input of power.
These decay mechanisms can be thought of as resulting in an energy confinement time $\tau_C$ for the confinement systems, leading to:
\begin{align}
	P_{C,e} &= U_C / \tau_C.
\end{align}
Of course, $\tau_C$ is likely not a constant function of $U_C$, but is likely to have many complex dependencies; but usually, power loss will increase with confined field energy, and so this is a useful parameter to consider.

For high-$Q_\text{eng}$ performance, the confinement power $P_{C,e}$ must be small compared to the output electrical power $P_\text{out} = \eta_B P_B + \eta_L P_L$.
Writing this in terms of $\beta$ and the confinement time, we have:
\begin{align}
	\tau_C \gg \frac{2}{3} \frac{1}{\beta \eta_L} \frac{\tau_E}{1 + \frac{\eta_B}{\eta_L} \frac{P_B}{P_L} }.
\end{align}
Thus, either the confinement system has to be very efficient relative to the plasma, in the sense of low power usage per field energy supplied, or the device has to operate at higher $\beta$.

\subsection{Electron Cyclotron Radiation}

So far, we have considered radiation from electrons due to bremsstrahlung.
However, in the presence of a magnetic field, electrons are also accelerated (and thus radiate), resulting in electron cyclotron (or synchrotron) radiation.

Owing to its complexity, the study of electron cyclotron radiation (ECR) has a long history \cite{Drummond1963ECR,trubnikov1979ecRadiation,Bornatici1983ECR,Kukushkin2008ECR}.
Much of this complexity comes from the fact that the plasma is optically thick to the emitted radiation, so that much of the power is reabsorbed.
Furthermore, in contrast to bremsstrahlung, this radiation tends to be lower-frequency, and thus can reflect from the surface of the confinement vessel.
Importantly, while these considerations make the study of ECR more complex, they also reduce the deleteriousness of the power loss.

While a quantitatively precise treatment of synchrotron radiation must be done for each specific system, a general estimation formulation has been developed, based on fitting to the results of many relevant simulations of the wave propagation \cite{Kukushkin2008ECR}.
Although only tested between 20 and 50 keV, the formula is nevertheless the best current way to get a rough estimate of the likely synchrotron power, and has been used in other studies of thermonuclear p-B11 fusion \cite{Cai2022TokamakPB11}.
Of course, a large degree of uncertainty is involved in extrapolating to these high temperatures.
With these caveats in mind, the formula for the effective power density is:
\begin{align}
	P_{EC} &\propto n_e^{1/2} T_e^{5/2} B^{5/2} (1-R_w)^{1/2} a^{-1/2} \lp 1 + 2.5 \frac{T_e}{E_\text{rest}} \rp,
\end{align}
where here, $a$ is the minor radius and $R_w$ is the wall reflection coefficient.

Consider the implication of this formula for a reactor design in terms of $\beta$:
\begin{align}
	P_{EC} &\propto n_e^{1.75} T_e^{3.75} \beta^{-1.25} (1-R_w)^{1/2} a^{-1/2} \lp 1 + 2.5 \frac{T_e}{E_\text{rest}} \rp.
\end{align}
Here, we see that for the purposes of reactor design, the EC power loss scales approximately as $n_e^2$, like the bremsstrahlung and fusion powers (at fixed ion mix).
We also see that it scales extremely strongly with temperature, as $T_e^{3.75}$.
Thus, for any magnetic confinement scheme, EC radiation will likely become an important part of the power balance.

The loss of power through EC emission is not necessarily such bad news for p-B11 fusion.
While bremsstrahlung power comes out in the form of hard x-rays, electrotron cyclotron power is lower frequency, and can often be reflected or absorbed.
This makes it much easier to envision achieving high power recycling efficiencies with EC power.

%\todo{I'm not 100\% sure that this is the best formula; we should investigate the high-temperature formulas referenced in Bornatici\cite{Bornatici1983ECR} more.}

\section{Conclusion} \label{sec:conclusion}

In this paper, we examined the power flow in an idealized p-B11 fusion reactor, and how alpha channeling could improve its performance.
As a performance metric, we calculated the required energy confinement time to achieve ignition, given the presence of strong bremsstrahlung radiation.
We showed that alpha channeling could bring down this required confinement time by an order of magnitude, and, as in \cite{kolmes2022waveSupported}, could be made even more effective by channeling energy into fast rather than thermal protons.
We also showed that channeling allowed the plasma to achieve ignition even in the presence of substantial contamination by ash.

One important caveat to our study is that the alpha channeling process is at this point only a theoretical possibility, since copious fusion-produced alpha particles are not yet produced even in DT experiments.  
However, alpha channeling relies only on the established, experimentally-validated quasilinear physics \cite{Kaufman1972,Eriksson1994} that underlies current drive \cite{Wong1980,Yamamoto1980,Kojima1981,Bernabei1982,Porkolab1984,Karney1985,fisch1987theory,Ekedahl1998} and, more directly, wave-induced ion transport from neutral beams \cite{Darrow1996EnhancedLoss,Fisch2000PhysicsAlpha}.
Furthermore, even though the mechanisms for alpha channeling are device-specific, there are theoretical studies that suggest that the channeling can happen in principle in a variety of devices, such as in tokamaks \cite{fisch1995ibw,Herrmann1997,Fisch1995a,White2021AlphaParticle,Castaldo2019,Cianfrani2019,Romanelli2020,ochs2015alpha,ochs2015coupling} or in mirror machines \cite{fetterman2008alpha,Fisch2006AlphaChanneling}.  
There have also been experiments in tokamaks that verified aspects of the relevant quasilinear theory \cite{Darrow1996EnhancedLoss,Fisch2000PhysicsAlpha}. 
Furthermore, because one of the main challenges of alpha channeling is to ensure that the timescale of wave-induced diffusion is fast than the collision timescale, the fact that the collision time for a fusion-born alpha particle in a p-B11 plasma ($\sim$ 1.1 s at $n_i = 10^{14}$ cm$^{-3}$, $T_i = 300$ keV, $T_e= 150$ keV) is more than twice as long as that of a fusion-born alpha particle in a DT plasma ($\sim$ 0.45 s at $n_i = 10^{14}$ cm$^{-3}$, $T_i = 20$ keV, $T_e= 20$ keV) means that similar levels of alpha channeling should be achievable with half the wave power in p-B11 plasmas.
Nevertheless, the alpha channeling effect remains speculative, as does a p-B11 reactor itself.  
What this study offers, however, is the recognition that, were alpha channeling successfully practiced, its effect on p-B11 ignition feasibility would be dramatic.

It must be acknowledged that even with alpha channeling, the required confinement times to achieve p-B11 fusion are formidable: around 66s for optimized, 100\% alpha channeling, at an ion density of $10^{14}$ cm$^{-3}$.
Furthermore, this is even before considering additional, device-specific loss mechanisms, such as the power required to support the confinement systems, and potential electron cyclotron radiation.

However, as discussed in Section~\ref{sec:recycling}, the situation is not necessarily so dire. 
Because the products of the fusion reaction are charged, much of the fusion power lost through thermal conduction can likely be recycled with high efficiency.
This would allow for lower confinement times, while still maintaining high $Q_\text{eng}$, even if ignition cannot be technologically achieved.
Such schemes would still require a large excess of fusion over bremsstrahlung power, as is provided by alpha channeling, but might be able to achieve this through additional means, such as deliberately deconfining high-energy electrons and recycling their energy.

Thus, even with the large improvement in feasibility presented in this paper, there remains a large space of possible high-Q configurations to explore.
The relative advantages of aneutronic fusion provide a strong incentive to explore this space more fully.

\section*{Acknowledgments}
We would like to thank Ariel S. Mitnick for helpful discussions on software engineering.
This work was supported by ARPA-E Grant DE-AR0001554. 
This work was also supported by the DOE Fusion Energy Sciences Postdoctoral Research Program, administered by the Oak Ridge Institute for Science and Education (ORISE) and managed by Oak Ridge Associated Universities (ORAU) under DOE contract No. DE-SC0014664.

\section*{Data Availability}

Data sharing is not applicable to this article as no new data were created or analyzed in this study.

\appendix

\section{Thermal Equilibration Rates} \label{app:equilibrationRates}

To calculate the collision frequencies, we primarily use the standard formulary formulae.
The power transfer rate coefficient for thermalization between Maxwellian species $a$ and $b$ is related to the thermalization collision frequency by $K_{ab} = 3 \nu_{ab} n_b / 2$.
For thermalization between protons and boron, we have:
\begin{align}
	K_{pb} &= 2.1 \times 10^{-7} \frac{Z_p^2 Z_b^2 m_p^{1/2} m_b^{1/2} \lambda_{pb}}{(m_p T_b + m_b T_p)^{3/2}} n_p n_b \text{ cm$^{-3}$s$^{-1}$}.
\end{align}

For collisions between fast ions ($f$, $\alpha$) with thermal ions ($p$, $b$), we have:
\begin{align}
	K_{fi} &= 2 K_{0fi} \lp \frac{m_f}{m_i} \phi(x_{fi}) - \phi'(x_{fi}) \rp,
\end{align}
where
\begin{align}
	K_{0fi} &= 9.0\times 10^{-8} Z_f^2 Z_i^2 \lambda_{ij} m_f^{1/2} E_f^{-3/2} n_f n_i \text{ cm$^{-3}$s$^{-1}$}\\
	x_{fi} &= \frac{m_i}{m_f}\frac{E_f}{T_i},
\end{align}
and $\phi(x_{fi})$ is a lower regularized incomplete gamma function: 
\begin{align}
	\phi(x_{fi}) = \frac{2}{\sqrt{\phi}} \int_0^{x_{fi}} t^{1/2} e^{-t} dt.
\end{align}

For thermalization between electron and ions ($p$, $b$), we have:
\begin{align}
	K_{ie} &= 4.8 \times 10^{-9} \frac{Z_i^2 \lambda_{ie}}{m_i T_e^{3/2}} n_i n_e R \text{ cm$^{-3}$s$^{-1}$}.
\end{align}
Here, $R$ is a relativistic correction factor from Putvinski\cite{Putvinski2019}:
\begin{align}
	R &= \frac{(1 + 2x + 2x^2) \sqrt{\pi x^3 / 2}}{\int_0^\infty t^2 e^{(-\sqrt{1+t^2} - 1)/x} dt},
\end{align}
where as in Eq.~(\ref{eq:PbrFormula}), $x = T_e / E_\text{rest}$.

Finally, for energy transfer from fast ions ($f$, $\alpha$) to electrons, we have:
\begin{align}
	K_{fe} &= 1.6 \times 10^{-9} Z_f^2 \lambda_{ef} n_f n_e R \text{ cm$^{-3}$s$^{-1}$} \notag\\
	&\qquad \times \lp 2 m_f^{-1} T_e^{-3/2} - 3 m_f^{-1} T_e^{-1/2} E_f^{-1} \rp .
\end{align}

\section{Fusion Cross Sections and Rates} \label{app:fusionRates}

The power balance calculations in this paper have depended on the fusion power production rate $P_F$. As was discussed in Section~III, this power output can be written as 
% will need to put in actual ref to section III's label rather than writing "III"
\begin{align}
	P_F = \mathcal{E}_F (K_{F,f} + K_{F,p}), 
\end{align}
where $\mathcal{E}_F = 8.7 \text{ MeV}$ is the energy produced per fusion reaction, $K_{F,f}$ is the rate of fusion reactions involving the fast proton population, and $K_{f,p}$ is the rate of fusion reactions involving the thermal proton population.

In general, the rate of fusion reactions between two species $s$ and $s'$ with distribution functions $f_s(\bv_s)$ and $f_{s'}(\bv_{s'})$ can be written as 
\begin{align}
	K_{F,ss'} = \int \D^3 \bv \, \D^3 \bv' \, \sigma(w) w f_s(\bv_s) f_{s'}(\bv_{s'}) , 
\end{align}
where $w \doteq |\bv_b - \bv_a|$ and $\sigma(w)$ is the cross-section for the fusion reaction between the two species. 

In this paper, we assume that the boron population is Maxwellian with temperature $T_b$ and density $n_b$: 
\begin{align}
	f_b = n_b \bigg( \frac{m_b}{2 \pi T_b} \bigg)^{3/2} \exp \bigg[ - \frac{m_b \bv_b^2}{2 T_b} \bigg], \label{eqn:maxwellianBoron}
\end{align}
where $m_b$ is the mass of a boron ion. 

For the fusion rate with the fast protons, the protons are taken to be a monoenergetic beam, so that 
\begin{align}
	\eta_\alpha(\bv_f) = \frac{n_f \delta(v_f - v_0)}{4 \pi v_0^2} \, ,
\end{align}
where $n_f$ is the fast-proton density, $v_f = |\bv_f|$, and $v_0$ can be written in terms of the beam energy $E_f$ as 
\begin{align}
	v_0 = \sqrt{ \frac{2 E_f}{m_p} } \, .
\end{align}
Here $m_p$ is the proton mass. 
Then the fast-proton fusion rate can be written as 
\begin{align}
	K_{F,f} = n_f \int \D^3 \bv_b \, \sigma(w) w f_b(\bv_b), 
\end{align}
where $\bw = \bv_b - v_0 \hat z$. Then 
\begin{align}
	K_{F,f} = n_f \int \D^3 \bv_b \, \sigma(w) w f_b(\bw + v_0 \hat z). 
\end{align}
Inserting Eq.~(\ref{eqn:maxwellianBoron}) for $f_b$, this is 
\begin{align}
	K_{F,f} 
	&= \frac{2 n_a n_b}{v_0} \bigg( \frac{m_b}{2 \pi T_b} \bigg)^{1/2} e^{- m_b v_0^2 / 2 T_b} \nonumber \\
	&\times \int_0^\infty \D w \, \sigma(w) w^2 \text{sinh} \bigg( \frac{m_b v_0 w}{T_b} \bigg) \exp \bigg[ - \frac{m_b w^2}{2 T_b} \bigg]. 
\end{align}
Define the reduced mass $\mu$ by $\mu \doteq m_p m_b / (m_p + m_b)$ and the center-of-mass energy $\varepsilon$ by 
\begin{align}
	\varepsilon \doteq \frac{\mu w^2}{2} . \label{eq:epsYBT}
\end{align}
Then in terms of $E_f$, 
\begin{align}
	&K_{F,f} = \frac{n_f n_b}{\mu^{3/2}} \bigg( \frac{2 m_p m_b}{\pi T_b E_f} \bigg)^{1/2} e^{- m_b E_f / m_p T_b} \nonumber \\
	&\times \int_0^\infty \D \varepsilon \, \sigma(\varepsilon) \varepsilon^{1/2} \, \text{sinh} \bigg( \sqrt{\frac{4 m_b^2}{m_p \mu}} \frac{E_f^{1/2} \varepsilon^{1/2}}{T_b} \bigg) \exp \bigg[ - \frac{m_b \varepsilon}{\mu T_b} \bigg]  
\end{align}
or, plugging in $m_b = 11 m_p$, 
\begin{align}
	&K_{F,f} = n_f n_b \bigg( \frac{3456}{121 m_p \pi T_b E_f} \bigg)^{1/2} e^{- 11 E_f / T_b} \nonumber \\
	&\times \int_0^\infty \D \varepsilon \, \sigma(\varepsilon) \varepsilon^{1/2} \, \text{sinh} \bigg( \frac{4 \sqrt{33} E_f^{1/2} \varepsilon^{1/2}}{T_b} \bigg) \exp \bigg[ - \frac{12 \varepsilon}{T_b} \bigg] . \label{eq:YBT}
\end{align}
Note that it is also possible to use $m_p$ in place of $\mu$ in Eqs.~(\ref{eq:epsYBT}-\ref{eq:YBT}), in which case one must use the cross section function $\sigma(w)$ with velocity $w$ defined in the boron rest frame, rather than the COM frame.

For $K_{F,p}$, the protons are instead assumed to be (approximately) Maxwellian with density $n_p$ and temperature $T_p$: 
\begin{align}
	f_p = n_p \bigg( \frac{m_p}{2 \pi T_p} \bigg)^{3/2} \exp \bigg[ - \frac{m_p \bv_p^2}{2 T_p} \bigg] . 
\end{align}
However, Putvinski suggested\cite{Putvinski2019} that modifications to the tail of $f_p$ would enhance the fusion yield. 
With that in mind, we approximate the yield by calculating the yield using a Maxwellian $f_p$ and then adding in an enhancement factor $\phi_k(T_p)$ to roughly match the results of Putvinski. 
$\phi_k$ is a piecewise linear function of $T_p$, going from $\phi_k(0 \text{ keV}) = 1.16$  to $\phi_k(700 \text{ keV}) = 1$, then $\phi_k(T_p > 700 \text{ keV}) = 1$ thereafter. 
With that in mind,  
\begin{align}
	K_{F,p} = \frac{2 n_p n_b \phi_k}{T^{3/2}} \sqrt{ \frac{2}{\pi \mu}} \int_0^\infty \D \varepsilon \, \sigma(\varepsilon) \varepsilon \exp \bigg[ - \frac{\varepsilon}{T} \bigg], 
\end{align}
where $\mu = (11/12) m_p$ and $T$ is the inverse-mass-weighted temperature 
\begin{align}
	T \doteq \frac{11 T_p + T_b}{12} \, ,
\end{align}
so that $K_{F,p}$ can also be written as 
\begin{align}
	K_{F,p} &= \frac{2 \cdot 12^{3/2} n_p n_b \phi_k}{(11 T_p + T_b)^{3/2}} \sqrt{ \frac{24}{11 \pi m_p}} \nonumber \\
	&\times \int_0^\infty \D \varepsilon \, \sigma(\varepsilon) \varepsilon \exp \bigg[ - \frac{12 \varepsilon}{11 T_p + T_b} \bigg]. 
\end{align}
The numerical calculations in this paper use interpolation between the data points reported by Sikora and Weller\cite{Sikora2016CrossSection} for the cross-section $\sigma(\varepsilon)$.

\section{Alpha Particle Collisional Power Transfer} \label{app:alphaCollisions}

Alpha particles are born with energies $E_{\alpha 0}$ according to a complicated distribution $S_\alpha(E_{\alpha 0})$ that depends on the energy distribution of the reactants\cite{Putvinski2019,Sikora2016CrossSection,Stave2011Understanding}.
Then, they transfer their energy collisionally to the various plasma constituents, until they roughly thermalize with the plasma around $E_\alpha \approx 3T_i / 2$, where $T_i \equiv (T_b + T_p)/2$.
During this process, the ratio $\alpha_{s0}$ of power transferred to each species is given by:
\begin{align}
	\alpha_{s0} = \frac{\int_{3T_i/2}^{\infty} dE_{\alpha 0}S_{\alpha} (E_{\alpha 0}) \int_{3T_i/2}^{E_{\alpha 0}} dE_\alpha \lp \frac{K_{\alpha s}}{\sum_{s'} K_{\alpha s'}} \rp}{\int_{3T_i/2}^{\infty} dE_{\alpha 0}S_{\alpha} (E_{\alpha 0}) \lp E_{\alpha 0} - 3T_i/2 \rp}.
\end{align}
Here, as discussed in Appendix~\ref{app:equilibrationRates}, the functions $K_{\alpha s}$ depend on $E_\alpha$, $T_s$, and $n_s$.
In addition, the distribution $S_\alpha(E_{\alpha 0})$ is a functional of $E_f$, $T_p$, $T_b$, $n_f$, $n_p$ and $n_b$.

Because a double integral is numerically expensive, we evaluate the integral at set points on a grid in parameter space, and then numerically (linearly) interpolate using python's scipy.ndimage.map\_coordinates() function.
To make this numerically efficient by reducing the number of dimensions, we make several key simplifications.

First, because the results are relatively insensitive to the starting alpha particle distribution, we simply use the shape of the uncorrected source distribution in Putvinski's\cite{Putvinski2019} Figure~B2, and then shift it by the typical reactant energy.
For the typical reactant energy, we use $3T_i / 2$.
%ignoring the slight corrections to the final result that would result from carefully integrating over the reactant energies and considering the beam; indeed, such corrections would have other effects on the distribution as well, since the angular cross sections of the various fusion channels are dependent on the reactant energies.
This approximation eliminates the dependence of $S_\alpha$ on $E_f$, reducing the dimensionality of the interpolating grid.

Second, rather than considering the boron and proton temperatures separately, we use the unweighted average temperature $T_i$ for each species, since the temperatures tend to be similar anyway (Fig.~\ref{fig:TVsEta}).
This approximation further reduces the dimensionality of the interpolating grid.

Finally, rather than treating the electron density as a separate parameter, we take the total ion density as $n_i = 10^{14}$ cm$^{-3}$, and take $n_e = Z_b n_b + (n_i - n_b)$. 
This approximation means that we miss a small amount of energy transfer to the electrons in Section~\ref{sec:ash}, since the alpha particles come with extra electrons that are not accounted for in the interpolator.
However, it does further reduce the dimensionality of the grid.

In sum, these approximations result in a fast interpolator for $\alpha_{s0}$ with four free parameters: $T_i$, $T_e$, $n_b$, and $f_p \equiv n_p / (n_p+n_f)$.
The speed of this interpolation is critical to the optimization, since optimizing for each set of parameters requires the evaluation of many power balance equilibria, and evaluating each power balance equilibrium requires calculating the derivatives $dU_s/dt$, which depend on $\alpha_{s0}$, many times.

%\clearpage\newpage

%\clearpage

%\bibliographystyle{unsrt}

%\bibliography{../../../Reading/allRefsZot.bib}
%\bibliography{comprehensive,More}

\clearpage

\end{document}